\input phyzzx
\newcount\mongocount
\mongocount=1
\def\Figure#1#2#3{
      \vbox to #3in{\hsize=#2in
        \vfil
         \includegraphics{#1}
    }
}
\def\figcap#1#2{
\vtop{\tenpoint\singlespace
\hsize=#1in\smallskip\noindent Figure\ \ \the\mongocount.\ \  #2
\global\advance\mongocount by 1\bigskip}}
\def\mongofigure#1#2#3#4#5{\centerline{\Figure{#1}{#2}{#3}
\figcap{#4}{#5}}}

\hoffset=0.375in
\overfullrule=0pt

\def\dol{{d_{\rm ol}}}
\def\dls{{d_{\rm ls}}}
\def\dos{{d_{\rm os}}}

\def\min{{\rm min}}

\def\kms{{\rm km}\,{\rm s}^{-1}}
\twelvepoint
\font\bigfont=cmr17
\centerline{\bigfont Measuring the Rotation Speed of Giant Stars}
\smallskip
\centerline{\bigfont From Gravitational Microlensing}
\bigskip
\centerline{{\bf Andrew Gould}\footnote{1}{Alfred P.\ Sloan Foundation Fellow}}
\smallskip
\centerline{Dept of Astronomy, Ohio State University, Columbus, OH 43210}
\smallskip
\centerline{gould@payne.mps.ohio-state.edu}
\bigskip
\centerline{\bf Abstract}
\singlespace 

	During some gravitational lensing events, the lens transits the
face of the star.  This causes a shift in the apparent radial velocity of
the star which is proportional to its rotation speed.  It also changes the
magnification relative to what would be expected for a point source.
By measuring both effects, one can determine the rotation parameter $v\sin i$.
The method is especially useful for K giant stars because these have turbulent 
velocities that are typically large compared with their rotation speed.  By
making a series of radial velocity measurements, one can typically determine
$v\sin i$ to the same accuracy as the individual radial velocity measurements.
There are approximately 10 microlensing transit events per year which would be 
suitable to make these measurements.

\bigskip
Subject Headings: gravitational lensing -- stars: rotation
\smallskip
\centerline{submitted to {\it The Astrophysical Journal}: 
November 7, 1996}
\smallskip
\centerline{Preprint: OSU-TA-29/96}

\endpage
\chapter{Introduction}

	There are two principal methods to determine the rotation speed $v$
of a star.  First, one can measure $v\sin i$ spectroscopically from the 
broadening of the spectral lines. Here $i$ is the angle between the line of 
sight and the spin axis of the star.  Second, one can measure the rotation
period $P$ photometrically from the periodic variation in luminosity due
to star spots.  Then from the known (or assumed) radius $r$, one finds the
rotation speed $v=2\pi r/P$.  However, neither method can be easily 
applied to giants.  Line broadening due to turbulence is typically 
$4-8\,\kms$ (Gray 1989).  Hence the small additional broadening 
due to expected rotations are difficult to detect.  For G giants, Gray (1989)
clearly detects $v\sin i$ of order $5\,\kms$ with errors of $\sim 1\,\kms$, but
K giants rotate more slowly and for these the detections are marginal.
To detect
rotation from star spots, the spots must remain stable during a time
$\gsim P$.  For giants with
$v\lsim 1\,\kms$ and $r\gsim 8\,r_\odot$, the rotation period is
$P\gsim 1\,$yr, which may be longer than the typical lifetime of spots.  In
any event, I know of no attempts to measure giant rotations from spots.

	Here I present an alternate method to measure the $v\sin i$ of giant 
stars
by making spectroscopic and photometric observations of a sub-class of 
ongoing microlensing events.  Microlensing occurs when the projected 
separation $\theta$ between a foreground object
(the lens) and a more distant source star is of order the
Einstein radius $\theta_e$,
$$\theta_e^2\equiv {4 G M \dls\over c^2 \dol\dos},\eqn\thetae$$  
where $M$ is the mass of the lens, and $\dol$, $\dls$, and $\dos$ are
the distances between the observer, lens, and source.
The lens then magnifies the source by an amount $A(x)$
$$A(x) = {x^2+2\over x(x^2+4)^{1/2}},\qquad x\equiv {\theta\over \theta_e}.
\eqn\aofx$$

	Let $\theta_*$ be the angular radius of the source.  If the lens
comes sufficiently close to the source, $\theta\lsim\theta_*$, the different
parts of the source will be magnified by significantly different amounts.
This differential magnification leads to two different kinds of effects.
First, the total magnification begins to deviate from the simple point
source formula $A(x)$.  This allows one to measure $x_*$,
$$x_* \equiv {\theta_*\over \theta_e},\eqn\rhodef$$
the stellar radius in units of the Einstein radius 
(Gould 1994; Nemiroff \& Wickramasinghe 1994; Witt \& Mao 1994).
Second, if the star is spinning, then the side which is moving toward us
will in general be closer to (or farther from) the lens than the
side which is moving away from us.  Since this side is magnified more, the
centroids of the stellar lines will be shifted to the blue (or the red).
From this shift one can measure the quantity,
$$U \equiv x_* v\sin i,\eqn\ldef$$
provided that the lens passes within a few stellar radii of the source
(Maoz \& Gould 1994).

	Heretofore, the main interest in these effects was that they allowed
measurement of $\theta_e$.  If $x_*$ is measured, and if the angular size of 
the source star is known
(as it usually is from Stefan's Law: flux $\propto T^4\theta_*^2$) 
then one can determine
$\theta_e = \theta_*/x_*$.  If both the angular size and the projected
rotation speed are known (as they might be for A stars observed in the Large
Magellanic Cloud) then $\theta_e = U\theta_*/v\sin i$.

	Here I point out that by combining photometric and spectroscopic
measurements, one can determine $x_*$ and $v\sin i$ separately.  I estimate
that it may be possible to use microlensing to determine $v\sin i$ for  
about 10 giants per year.

\chapter{Microlensing at Close Quarters}

	Three groups have been monitoring a total of several $10^7$ stars
toward the Galactic bulge (Alcock et al.\ 1995; Udalski et al.\ 1994;
Alard 1995).  Together they have observed more than $100$ microlensing events.
A fourth group (Aubourg et al.\ 1993; Ansari 1996) has recently joined 
this search. 
 If these
searches are tuned primarily to finding lensing events of bulge giants, then
events could be found at a rate $\sim 170\,\rm y^{-1}$ including
$\sim 10\,\rm yr^{-1}$ in which the lens transits the face fo the star,
i.e., events where $\theta<\theta_*$ (or $x<x_*$) at the peak (Gould 1995b).  
It is these
transit events that provide the main opportunity to measure $v \sin i$. 

\FIG\one{
Deviation from point-source behavior as a function of $z$, the 
lens-source separation in units of the source size.  Solid curve is the
magnification adjustment factor $B_0(z)$ which should be multiplied by the 
naive point-source magnification $A(x)$ to give the true magnification.
Bold curve is (four times) the velocity shift factor $G(z)$.  The apparent
shift in radial velocity is given by $v\sin i \sin \gamma G(z)$, 
where $v\sin i$
is the projected rotation speed and $\gamma$ is the angle between the projected
source axis and the position of the lens.  Note that the two functions track
one another very well for $z\lsim 1$.
}
\topinsert
\mongofigure{ps.spin}{6.4}{5.5}{6.4}
{
Deviation from point-source behavior as a function of $z$, the 
lens-source separation in units of the source size.  Solid curve is the
magnification adjustment factor $B_0(z)$ which should be multiplied by the 
naive point-source magnification $A(x)$ to give the true magnification.
Bold curve is (four times) the velocity shift factor $G(z)$.  The apparent
shift in radial velocity is given by $v\sin i \sin \gamma G(z)$, 
where $v\sin i$
is the projected rotation speed and $\gamma$ is the angle between the projected
source axis and the position of the lens.  Note that the two functions track
one another very well for $z\lsim 1$.
}
\endinsert

	The finite size of the source causes the light curve to deviate
from its standard form according to,
$$A(x)\rightarrow A(x)B_0(z),\qquad z\equiv{\theta\over \theta_*},\eqn\abofxz$$
where $x$ is now regarded as the separation between the lens and the
{\it center} of the star and $B_0(z)$ is the function shown in Figure \one\
(Gould 1994).  Because of the structure in this curve at $z\sim 1$, it is 
quite easy to measure $x_*$ whenever $z$ reaches a minimum value 
$z_\min\lsim 1$.
[Note from Fig.\ \one\ that the deviation is significant even for $z\sim 2$.
However, a more detailed analysis shows that with
single-band photometry it is not possible to measure $x_*$
unless $z\lsim 1$.  Such measurements are possible out to $z\sim 2$
using two-band optical/infrared photometry (Gould \& Welch 1996).]

	In computing $B_0(z)$ I have assumed that the star is limb-darkened,
with surface brightness,
$$S(z) = S(0)[ 1 - \kappa_1 Y - \kappa_2 Y^2],\qquad Y\equiv 1-(1-z^2)^{1/2}
\eqn\limbeq$$
and have adopted parameters $\kappa_1=0.567$ and $\kappa_2=0.114$ suitable
for a star at $T=4500$ K observed in I band.  As I discuss in \S\ 4,
uncertainties in the limb-darkening coefficients play a very small role.

	The finite size of the source also induces a shift in the center
of the spectral lines.  Let $\gamma$ be the angle between the projected
axis of source rotation and the position of the lens.  Then the line shift
is given by
$$\Delta v(z,\gamma) = 
{\int_0^1 d z'\,z' S(z')\int_0^{2\pi}d\phi\, A(q x_*) (z'\sin\phi v\sin i) 
\over
 \int_0^1 d z'\,z' S(z')\int_0^{2\pi}d\phi\, A(q x_*)},\eqn\delvone$$
where $(z',\phi)$ is the position on the star relative to the projected
spin axis and
$q^2\equiv {z^2 + z'^2 - 2 z z'\cos(\phi-\gamma)}$.
Assuming $x_*\ll 1$ this becomes,
$$\Delta v(z,\gamma) = v\sin i\sin\gamma\, G(z),\qquad
G(z)\equiv {B_1(z)\over B_0(z)},\eqn\delvtwo$$
where
$$B_n\equiv 
{\int_0^1 d z'\,z' S(z')\int_0^{2\pi}d\psi 
[1 + (z'/z)^2 - 2(z'/z)\cos\psi]^{-1/2} (z'\cos\psi)^n
\over 2\pi \int_0^1 d z'\,z' S(z')},\eqn\bsubn$$
and $\psi=\gamma-\phi$.
The function $G(z)$ is shown in Figure \one.

	The source-lens separation is given by,
$$x(t) = [\omega^2(t-t_0)^2 + \beta^2]^{1/2},\eqn\xoft$$
where $t_0$ is the time of maximum magnification, $\beta$ is the impact
parameter in units of $\theta_e$ and $\omega^{-1}$ is the Einstein-radius
crossing time.  
Let $\alpha$ be the angle between 
the lens direction of motion and the projected spin axis of the source.
Then equation \delvtwo\ can be written
$$\Delta v(t) = v\sin i 
{\beta\cos\alpha + \omega (t-t_0)\sin \alpha\over x(t)}
G\biggl[{x(t)\over x_*}\biggr]
.\eqn\delvfive$$
For giant-star events with low impact parameter $(\beta\ll 1)$, the parameters
$t_0,\ \beta,$ and $\omega$ are usually determined to $<1\%$ from the
overall light curve, implying that $x(t)$ is also well 
determined.  Hence, equation \delvfive\ is effectively a function of
three parameters, $v\sin i$, $\alpha$, and $ x_*$.  In principle, one
could fit the time series of observed line shifts for these three
parameters.  In practice, for events with $\beta\lsim x_*$, it is much easier
to measure $ x_*$ by fitting photometric measurements to 
$A(x)B_0[x(t)/ x_*]$.  Hence, $ x_*$ may also be regarded as a known quantity
in equation \delvfive, implying that $\Delta v$ is effectively a function of 
just two parameters, $v\sin i$ and $\alpha$.

	To make contact with the work of Maoz \& Gould (1994), I note
that for $\beta\gsim x_*$, $G(z) \rightarrow \xi/(4z)$ where
$\xi = 2\int dz z^3 S(z)/\int d z z S(z)$, with $\xi\sim 0.89$ for $I$ band.
This means that $\Delta v \propto  x_* v\sin i$, so that it is impossible
to find $ x_*$ and $v\sin i$ separately.  Hence, unless the lens transits
(or nearly transits) the source, or there is some other information about
$ x_*$ (see \S\ 4), one cannot use spectroscopy of lensing events to 
determine $v\sin i$.

	Equation \delvfive\ was obtained under the assumptions $ x_*\ll 1$
and $x\ll 1$.  If one drops the latter assumption and repeats the derivation
beginning with equation \delvone, one finds that equation \delvfive\ remains
valid provided that one makes the substitution,
$$G\biggl({x\over  x_*}\biggr)\rightarrow \tilde G(x, x_*)\equiv
G\biggl({x\over  x_*}\biggr)\biggl[-x{d\ln A\over x}\biggr]
={G(x/ x_*)\over (1 + x^2/2)(1+x^2/4)}.\eqn\delvsix$$

\chapter{Error Analysis}

	Suppose that a series of measurements $u_k$ are made of the apparent
radial velocity of the star at times $t_k$, with errors $\sigma_k$.  One
then fits the measurements to a function of the form,
$$F(t;v_0,\eta,\alpha,v\sin i) = v_0 + \eta t + \Delta v(\alpha,v\sin i),
\eqn\fdef$$
where $v_0$ is the true velocity of the source at the peak of the event,
$\eta$ is the radial acceleration of the source, and $\Delta v(\alpha,v\sin i)$
is given by equation \delvfive.  Then one may estimate the covariances 
$c_{i j}$
of the determinations of the parameters 
$(a_1,a_2,a_3,a_4) = (v_0,\eta,\alpha,v\sin i)$ by (e.g.\ Gould 1995a),
$$c= b^{-1},\qquad b_{i j} = \sum_k \sigma_k^{-2}
{\partial F(t_k)\over \partial a_i}
\,{\partial F(t_k)\over \partial a_i}.\eqn\candb$$
Of primary interest is $c_{4,4}^{1/2}$, the error in $v\sin i$.

\midinsert
\nopagenumbers
$$
\vbox{\halign{#\hfil\quad&
\hfil#\hfil\quad&\hfil#\hfil\quad&\hfil#\hfil\quad&\hfil#\hfil\cr
\multispan5{\hfil TABLE 1 \hfil}\cr
\noalign{\smallskip}
\multispan5{\hfil `Three Observatories'\hfil}\cr
\noalign{\medskip}
\multispan5{\hfil Errors in $v\,\sin i$\hfil}\cr
\noalign{\smallskip}
\multispan5{\hfil (In units of measurement error, $\delta v$)\hfil}\cr
\noalign{\smallskip}
\noalign{\hrule}
\noalign{\smallskip}
\noalign{\hrule}
\noalign{\smallskip}
&\multispan4{\hfil Position Angle\hfil}\cr
&\multispan4{\hfil Relative to Spin Axis\hfil}\cr
\noalign{\vskip-9pt}
&\multispan4{\hrulefill}\cr
\noalign{\smallskip}
$\beta/x_*$&
$\alpha= 0^\circ$& $\alpha=30^\circ$& $\alpha=60^\circ$& $\alpha=90^\circ$\cr
 
    0.2&   4.6&   3.9&   2.2&   0.9\cr
    0.4&   2.3&   1.9&   1.2&   1.0\cr
    0.6&   1.6&   1.4&   1.1&   1.1\cr
    0.8&   1.4&   1.2&   1.3&   1.4\cr
    1.0&   1.7&   1.5&   1.6&   1.9\cr
    1.2&   2.3&   2.0&   2.0&   2.4\cr
    1.4&   3.0&   2.4&   2.4&   3.0\cr
\cr
\noalign{\smallskip}
\noalign{\hrule}
}}
$$

\endinsert

	In general, the errors will depend on the details of the experimental
setup and on the observing conditions.  In order illustrate the overall
sensitivity of the observations I adopt a specific model.  I assume that
measurements are made at a rate $20\omega/ x_*$, that is 20 times per
source-radius crossing time.  For typical events seen toward the bulge,
$ x_*/\omega\sim 10$ hours $(r_s/10\,r_\odot)$ where $r_s/r_\odot$ is the
radius of the source in solar units.  I define $\delta v$ as the error
at the peak of the event and assume that for other measurements, the
errors scale inversely as the square root of magnification.  That is,
$\sigma_i[x(t_i)] = \delta v[B_0(\beta/ x_*)x/B_0(x/ x_*)\beta]^{1/2}$.
I assume that the measurements are carried out from two source crossing times
($2 x_*/\omega$) before the peak until 5 crossing times after the peak.
The results expressed in units of $\delta v$ are shown in Table 1 for 
various values of $\alpha$ and $\beta/ x_*$ (impact parameter in units of 
$\theta_*$).  I have not shown the results for $\alpha>90^\circ$ which
are almost exactly equal to the results for $180^\circ-\alpha$.

\midinsert
\nopagenumbers
$$\vbox{\halign{#\hfil\quad&
\hfil#\hfil\quad&\hfil#\hfil\quad&\hfil#\hfil\quad&\hfil#\hfil\cr
\multispan5{\hfil TABLE 2 \hfil}\cr
\noalign{\smallskip}
\multispan5{\hfil `Two Observatories'\hfil}\cr
\noalign{\medskip}
\multispan5{\hfil Errors in $v\,\sin i$\hfil}\cr
\noalign{\smallskip}
\multispan5{\hfil (In units of measurement error, $\delta v$)\hfil}\cr
\noalign{\smallskip}
\noalign{\hrule}
\noalign{\smallskip}
\noalign{\hrule}
\noalign{\smallskip}
&\multispan4{\hfil Position Angle\hfil}\cr
&\multispan4{\hfil Relative to Spin Axis\hfil}\cr
\noalign{\vskip-9pt}
&\multispan4{\hrulefill}\cr
\noalign{\smallskip}
$\beta/x_*$&
$\alpha= 0^\circ$& $\alpha=30^\circ$& $\alpha=60^\circ$& $\alpha=90^\circ$\cr
     0.2&   4.6&   4.2&   2.8&   1.3\cr
    0.4&   2.3&   2.3&   1.9&   1.4\cr
    0.6&   1.6&   1.8&   1.8&   1.6\cr
    0.8&   1.4&   1.7&   1.9&   1.9\cr
    1.0&   1.7&   1.9&   2.2&   2.2\cr
    1.2&   2.3&   2.3&   2.4&   2.5\cr
    1.4&   2.9&   2.7&   2.7&   2.9\cr

\cr
\noalign{\smallskip}
\noalign{\hrule}
}}
$$

\endinsert

\midinsert
\nopagenumbers
$$\vbox{\halign{#\hfil\quad&
\hfil#\hfil\quad&\hfil#\hfil\quad&\hfil#\hfil\quad&\hfil#\hfil\cr
\multispan5{\hfil TABLE 3 \hfil}\cr
\noalign{\smallskip}
\multispan5{\hfil `One Observatory'\hfil}\cr
\noalign{\medskip}
\multispan5{\hfil Errors in $v\,\sin i$\hfil}\cr
\noalign{\smallskip}
\multispan5{\hfil (In units of measurement error, $\delta v$)\hfil}\cr
\noalign{\smallskip}
\noalign{\hrule}
\noalign{\smallskip}
\noalign{\hrule}
\noalign{\smallskip}
&\multispan4{\hfil Position Angle\hfil}\cr
&\multispan4{\hfil Relative to Spin Axis\hfil}\cr
\noalign{\vskip-9pt}
&\multispan4{\hrulefill}\cr
\noalign{\smallskip}
$\beta/x_*$&
$\alpha= 0^\circ$& $\alpha=30^\circ$& $\alpha=60^\circ$& $\alpha=90^\circ$\cr
 
    0.2&  14.3&  12.0&   6.4&   1.4\cr
    0.4&   6.4&   5.1&   2.6&   1.3\cr
    0.6&   3.7&   2.9&   1.6&   1.4\cr
    0.8&   2.7&   2.1&   1.5&   1.8\cr
    1.0&   3.2&   2.5&   2.1&   2.5\cr
    1.2&   5.1&   3.8&   2.8&   3.5\cr
    1.4&   7.1&   5.2&   3.7&   4.8\cr

\cr
\noalign{\smallskip}
\noalign{\hrule}
}}
$$

\endinsert

	From Table 1, one sees that, for example, with radial velocity errors 
of $\delta v\sim 0.1\,\kms$ one
could typically expect to measured $v\sin i$ to an accuracy of 0.2--0.3
$\kms$, although the errors are larger for some unfavorable geometries.

	Next, I examine how critically these results depend on the
continuous coverage assumed in the construction of Table 1.  In Table 2
I have assumed no observations are made during every third crossing
time.  In Table 3, I assume no observations are made for two out of
every three crossing times.  The three tables then roughly approximate
what could be done with 3, 2, and 1 observatories, respectively.
In order to facilitate comparison with
Table 1, I assume the same number of total observations.  That
is observations are carried out at a rate $30 \omega/x_*$ for Table 2
and at $60 \omega/x_*$ for Table 3.  It is clear from these tables
that organizing observations from several sites around the world
has important advantages.

	One possible method of further reducing the errors is to obtain
independent information on the zero-point velocity $v_0$ and the
acceleration $\eta$.  Such constraints could in principle be obtained
by observing the source after the event was over to look for periodic
motion due to a companion.  Unfortunately, as I now show, obtaining
significant improvements on the errors reported above would be quite difficult.
By examining the covariance matrix $c_{i j}$, I find that typically
$(c_{2,2}/c_{4,4})^{1/2}\sim 0.1\omega/ x_*$.  This means that if, for
example, $v\sin i$ were determined with accuracy 300 m s${}^{-1}$, and
the crossing time were $\sim 0.5\,$day, then $\eta$ would be known
with an accuracy of 60 m s${}^{-1}$ day${}^{-1}$, about 1/8 the acceleration
of the Earth.  To obtain a significant improvement on the measurement of
$v\sin i$ one would have to measure $\eta$ independently to at least
this accuracy.  However, such a small acceleration could be produced
by, for example, a planet of 4 Jupiter masses at 0.2 AU.  Such a planet
would generate redshift 
oscillations with an amplitude of only 300 m s${}^{-1}$.
These oscillations would be extremely difficult to detect.  The main
reason that one can hope to measure subtle changes in apparent redshift
during the lensing event is that the star is magnified $\gsim 10$ times.
After the event is over and the star returns to its unlensed luminosity,
typically $I\lsim 16$, highly accurate velocity measurements become
extremely difficult.

\chapter{Discussion}

	In the analysis given in \S\ 2, I implicitly assumed that the
local strength of the lines used to measure the redshift is proportional
to the local surface brightness.  In fact, since limb-darkening is caused
in part by line absorption, one expects that the line strength is
more heavily weighted toward the limb of the star than the luminosity.
This poses no fundamental difficulties since one can simply 
replace $S(z)$ with the line-strength density in equation \delvone\ and 
following.  The two are calculated in the same stellar-atmosphere codes.
However, the complication does break the purely empirical link between
the determination of $ x_*$ from photometry (which does depend only
on $S$), and the use of $ x_*$ in the interpretation of the line-shift
data.  This link is illustrated by the similar forms of $B_0$ and $G$
for $z\lsim 1$ in Figure \one.
For most cases the systematic errors induced by this effect will be small.  

	I now briefly describe several possible extensions of the basic 
method presented in this paper.  The most straight forward extension is
to use Ca II line at 393.3 nm to measure the line shift.  
Loeb \& Sasselov (1995) showed that narrow-band imaging of this line during
a lensing event could be used to measure $ x_*$.  Giant stars are
limb-{\it brightened} in Ca II, so the surface brightness has a ring-like
structure.  The same feature would make it especially useful for monitoring
line shifts during the event.  The formalism presented in \S\ 2 is easily
adapted to this case simply by replacing $S(z)$ with the emission profile.
If one used the Loeb \& Sasselov (1995) photometric method for determining
$ x_*$ (either by measuring the absolute line fluxes from spectroscopy
or by independent narrow-band imaging) then systematic errors would
be small.  On the other hand, if $ x_*$ were measured by some other means,
then the accuracy of the results would depend more critically on the
accuracy of the model.  A limitation of this method is that many
bulge fields are heavily reddened.  Since the extinction at 393 nm is
$\sim 2.7$ times greater than at $I$, it will often be much easier
to measure redshifts by cross-correlation in the $I$ band than from the Ca II
line.

	If the stellar radius could be measured for $ z\gsim 1$, then
$v\sin i$ could be determined for a much larger number of giants.  This
is because for $z\gsim 1$, the line shift falls off only as
$\Delta v\propto z^{-1}$ (Maoz \& Gould 1995).  One method of extending
the measurement to $ x_*\sim 2$ is optical/infrared photometry.  
In this method, the inferred value of $ z$ is
$\propto (\Lambda^H - \Lambda^V)^{-1}$, where $\Lambda$ is proportional
to the second moment of the surface brightness in $V$ and $H$ (Gould \& Welch
1996).  Since the modeling of this combination of parameters is relatively
independent of the modeling of the line-strength profile, there is a
somewhat greater possibility of systematic errors.  

	Finally, it may be possible to use optical
interferometry to directly image the two source images during a lensing
event (Gould 1996).  
This method does not depend at all on the lens being near the face of
the source.  Direct imaging yields a measurement of $\theta_e$ and also of the
direction of motion (Gould 1992).  
If $\theta_*$ can be determined from Stefan's Law,
then $ x_*=\theta_*/\theta_e$ is determined.  The image separation
is $2\theta_e$, which typically is of order $100\mu$as.  A highly magnified
giant might be $I\sim 14$.  It will be at least a few years before
such measurements are possible, but this method may permit the measurement
of $v\sin i$ for a large number of giants.  Recall that the line-shifts
also measure $\alpha$, the angle between the projected spin axis and the
the direction of lens motion.  Since optical interferometry measures
the absolute direction of lens motion, one also obtains the absolute
direction of the projected spin axis.

{\bf Acknowledgements}:  I would like to thank M.\ Pinsonneault for
stimulating discussions.  This work was supported in part by grant 
AST 94-20746 from the NSF.

\endpage
\Ref\alard{Alard, C.\ 1996, in IAU Symp.\ 173 ed.\ C.\ S.\ Kochanek \&
J.\ N.\ Hewitt) (Dordrecht: Kluwer), 215}
\Ref\Alcock{Alcock, C., et al.\ 1995, ApJ, 445, 133}
\Ref\Ansari{Ansari, R., et al.\ 1996, A\&A, in press}
\Ref\Aubourg{Aubourg, E., et al.\ 1993, Nature, 365, 623}
\Ref\gtwo{Gould, A.\ 1992, ApJ, 392, 442}
\Ref\gtwo{Gould, A.\ 1994, ApJ, 421, L71}
\Ref\gthree{Gould, A.\ 1995a, ApJ, 440, 510}
\Ref\gthree{Gould, A.\ 1995b, ApJ, 447, 491}
\Ref\gthree{Gould, A.\ 1996, PASP, 108, 465}
\Ref\gw{Gould, A.\ \& Welch, D.\ L.\ 1996, ApJ, 464, 212}
\Ref\gray{Gray, D.\ F.\ 1989, ApJ, 347, 1021}
\Ref\ls{Loeb, A.\ \& Sasselov, D.\ 1995, ApJ, 449, L33}
\Ref\mg{Maoz, D.\ \& Gould, A.\ 1994, ApJ, 425, L67}
\Ref\nem{Nemiroff, R.\ J.\ \& Wickramasinghe, W.\ A.\ D.\ T.\ ApJ, 424, L21}
\Ref\oglea{Udalski, A., et al.\ 
1994, Acta Astronomica 44, 165}
\Ref\wittmao{Witt, H.\ \& Mao, S.\ 1994, ApJ, 430, 505}

\refout
\endpage
\bye